\pdfoutput=1

\documentclass[aps,prl,twocolumn,superscriptaddress,showpacs,floatfix]{revtex4}
\usepackage{graphicx}

\begin{document}

%\preprint{}

\title{Correlated electrons in optically-tunable quantum dots:
\\ Building an electron dimer molecule}

\author{Achintya Singha}
\altaffiliation{Present address: Dept of Physics, Bose Institute, 93/1 Acharya Prafulla Chandra Road, Kolkata 700009, India.}
\author{Vittorio Pellegrini}
\email{vp@sns.it}
\affiliation{NEST, Istituto di Nanoscienze -- CNR and Scuola Normale Superiore, Pisa 56127, Italy}
%\affiliation{NEST CNR-INFM and Scuola Normale Superiore, Pisa 56126, Italy}
\author{Aron Pinczuk}
\affiliation{Depts of Appl. Phys \& Appl. Math. and of Physics, Columbia University, New York 10027, USA}
\author{Loren N. Pfeiffer}
%\affiliation{Bell Laboratories, Alcatel-Lucent, Murray Hill, New Jersey 07974, USA}
\author{Ken W. West}
\affiliation{Department of Electrical Engineering,
Princeton University, Princeton, NJ, USA and Bell Laboratories, Alcatel-Lucent, Murray Hill, New Jersey 07974, USA}
\author{Massimo Rontani}
\email{rontani@unimore.it}
\affiliation{S3, Istituto di Nanoscienze -- CNR, Modena 41125, Italy}


%\author{}
%\thanks{}
%\affiliation{}


\date{\today}

\begin{abstract}
We observe the low-lying excitations of a molecular dimer 
formed by two electrons in a GaAs semiconductor 
quantum dot in which 
the number of confined electrons is tuned by optical illumination.
By employing inelastic light scattering we identify the inter-shell 
excitations in the one-electron regime and the distinct spin and charge 
modes in the interacting few-body configuration.
In the case of two electrons a comparison with configuration-interaction 
calculations allows us to link the observed excitations with the 
breathing mode of the molecular dimer and to determine the singlet-triplet 
energy splitting.
\end{abstract}


\maketitle
\par
The creation and control of electrons confined in semiconductor quantum
dots (QDs) enables fundamental studies of Coulomb
interaction at the nanoscale and allows to define optimal architectures
for solid-state quantum information processing \cite{hanson07}.
In the case of two electrons, singlet and triplet spin states with a 
well-defined energy splitting are under active investigation for the 
implementation of logic gates for quantum 
computing \cite{burkard,marcus04,hanson07}.
\par
In the dilute limit of few electrons, a paradigm shift occurs as the
liquid-like properties of the electrons --captured by a
single-particle picture-- cease to dominate and the impact of Coulomb
correlation favors the formation of a molecular electron state
with strong radial and angular order
\cite{bryant, maksym, egger, reimann, ghosal, kalliakos08}. The molecular
dimer formed by two electrons is especially relevant being the benchmark
for the exploration
of Coulomb correlation at the nanoscale. This state has been
the subject  of several theoretical investigations
(e.g. \cite{yannouleas00, matulis01, pfannkuche, Rontani99}) albeit difficult 
to realize experimentally since it requires soft
confinement potentials that are comparable to Coulomb interaction strengths.
Two-electron correlated QDs were studied
in magneto-transport experiments at finite fields
\cite{ellenberger06}.
\par
Like in atomic nuclei \cite{bohr}, the emergence of strong correlation
among the QD electrons manifests in the energetic structure of the
low-lying few-body modes that are accessible by inelastic light 
scattering \cite{garcia,kalliakos08}.
Optical investigations of the characteristic low-lying excitations of 
two electrons in the QD remain challenging. 
Such experiments are enabled by combining spectroscopy tools with 
charge tunability
with single electron accuracy, as obtained in devices containing
self-assembled InAs QDs \cite{drexler94, broke03} where two-electron 
singlet and triplet
states were observed by inelastic light
scattering \cite{heitmann09}. In these self-assembled QDs, however, the large
confinement energy prevents the formation of a molecular dimer state.
\par
In this Letter we report the identification of a dimer molecular 
state of two electrons at zero magnetic field
in nanofabricated GaAs/AlGaAs QDs. The evidence is found in the 
low-lying spin and charge excitations
observed by resonant inelastic light scattering. Crucial to these 
experiments is the
continuous charge tunability by photo-depletion that allows 
the creation of QDs with a selected
number of electrons. We first identify the one-electron regime 
characterized by a single monopole excitation
between Fock-Darwin (FD) levels ---the states of the
two-dimensional harmonic oscillator-- that are simply 
linked to the in-plane confinement energy.
\par
The observation of additional distinct spin and charge modes as electron 
numbers in the QDs are increased as we sweep the illumination intensity 
signals the emergence of a ground state composed by two and more 
interacting electrons.
We show that the measured low-lying excitation mode energies are 
incompatible with a theoretical framework that neglects correlation 
effects. We propose an interpretation based on a configuration 
interaction (CI) approach (aka exact diagonalization) that precisely 
reproduces the inelastic light scattering spectra. 
In the case of two electrons the CI analysis reveals links between 
observed excitations and the breathing and center-of-mass
oscillations and rigid-rotor rotations of the molecular electron dimer.
\par
In the CI interpretation the rotation energies are 
$\hbar^2M^2_{\text{rel}}/ 2I$, 
with $I$ being the momentum of inertia and 
$M_{\text{rel}}=0,\pm 1,\pm2, \ldots$ the angular momentum in the 
relative frame. To comply with the
antisymmetry of the total two-electron wave function, $M_{\text{rel}}$ must
be even (odd) for a singlet (triplet) state. Therefore, the energy
separation between the triplet and singlet ground states is the
rotation quantum $\hbar^2/2I$, that as shown below is accessible by
the inelastic light scattering experiments.
\par
\begin{figure}
\begin{center}
\includegraphics*[width=8.5cm]{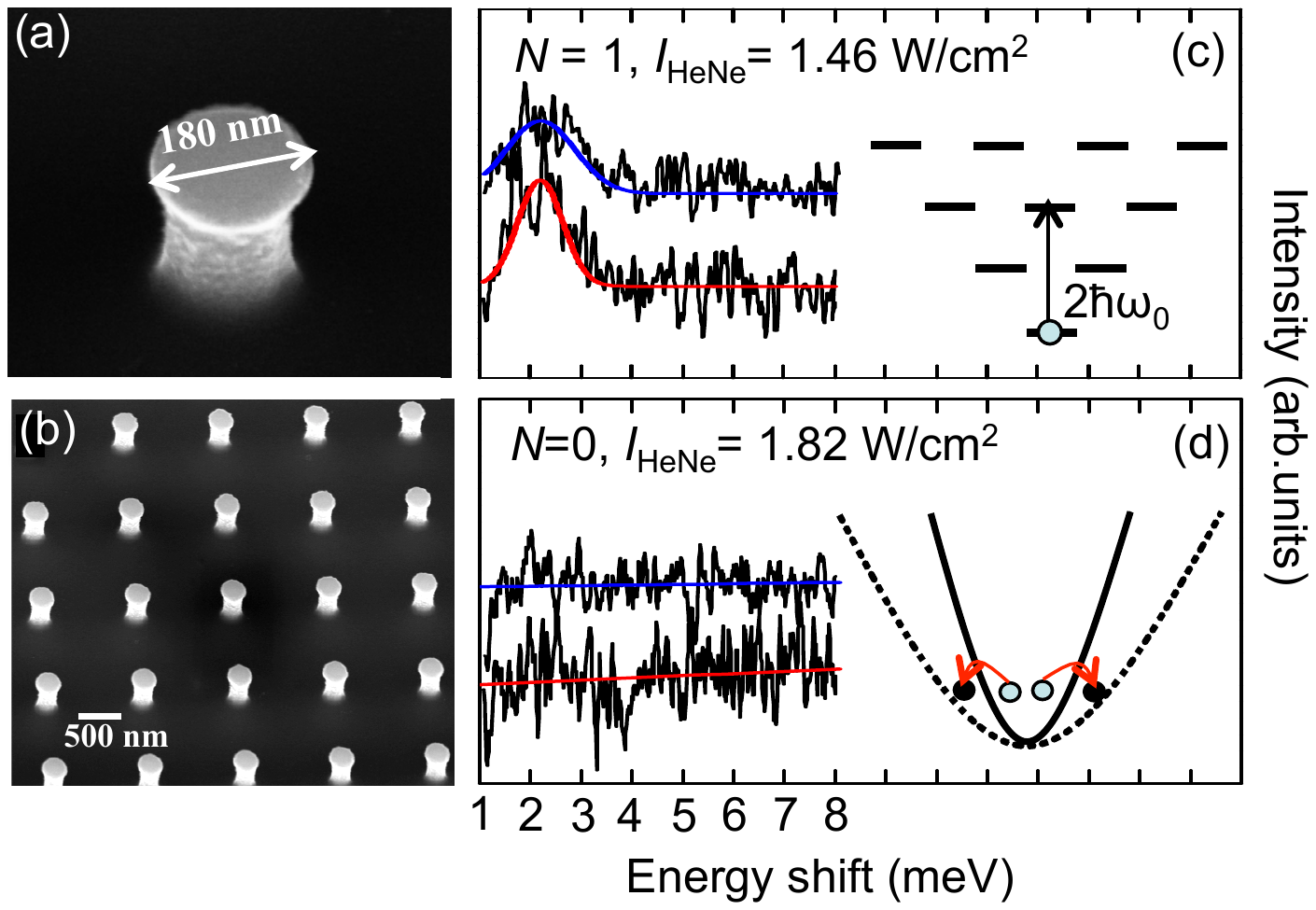}
\end{center}
\caption{(Color)  (a) SEM images of a QD pillar and
(b) of the QD array. (c) and (d) Resonant inelastic light scattering spectra
of spin (fitted with red line) and charge (blue line) excitations
for $N=0$ and $N=1$. The $N=1$
intershell transition is shown in the inset to (c).
The horizontal segments are
the FD levels. The sketch of the
QD in-plane parabolic potential is shown in the inset to (d).
The photo-generated holes (filled black circles)
accumulate in the depletion region close to the
physical borders of the etched pillar provoking a flattening of the
confining potential (dotted curve). Removal of the electrons
(filled light-blue circles) occurs via photo-recombination with
the photo-excited holes (red arrows).\label{fig1}}
\end{figure}

The sample is constructed as an array (100 $\mu$m  $\times$ 100 $\mu$m) 
of identical
QDs ($10^4$ replica to improve the signal-to-noise ratio displaying the same
optical properties \cite{gurioli} and inter-dot spacing of 1 $\mu$m).
It is fabricated by electron beam lithography and inductively
coupled-plasma reactive ion etching on a 25 nm wide, one-side
modulation-doped Al$_{0.1}$Ga$_{0.9}$As/GaAs quantum well. The measured
low-temperature two-dimensional electron density and mobility are
$1.1 \cdot 10^{11}$ cm$^{-2}$ and $2.7 \cdot 10^6$ cm$^2$/Vs,
respectively. SEM images of such QDs are shown
in Figs.~\ref{fig1}(a) and (b).
\par
The inelastic light scattering experiments were
performed in a backscattering configuration ($q \le 2 \cdot 10^4$ cm$^{-1}$
where $q$ is the wave-vector transferred into the lateral dimension) at
$T = 2$ K. A tunable ring-etalon Ti:Sapphire laser with a wavelength of
789 nm in resonance with the QD absorption (whose onset
is at 813 nm) was focused on the QD array with a 100 $\mu$m diameter
area, and the scattered light was collected into a triple grating
spectrometer with CCD detection.
\par
Photo-depletion of the QD electrons was achieved by continuous-wave
HeNe laser excitation at 633 nm that creates electron-hole pairs
in the AlGaAs barrier. Whereas the photoexcited electrons in the AlGaAs
barrier contribute to charge compensation of the ionized donors, the
photoexcited holes are swept in the QDs flattening the parabolic
in-plane potential (i.e. decreasing the confinement energy $\hbar \omega_0$
and contributing to the removal of the confined
electrons by electron-hole recombination \cite{kuku89,kalliakos08nano})
[cf.~inset to Fig.~\ref{fig1}(d)].
As shown below the method allows to fully photo-deplete the
QDs and tune their population with single-electron accuracy by changing
the HeNe laser intensity $I_{\text{HeNe}}$.
\par
\par
Distinct spin and charge intershell monopole modes
(with $\Delta M = 0$, $M$ is the total angular momentum) were detected as a
function of $I_{\text{HeNe}}$ by varying the linear polarization of
the scattered photons [parallel (perpendicular) to that of the incoming
laser for charge (spin)]. The condition of photo-depleted QDs (number of
confined electrons $N=0$) is defined by the absence of any inter-shell
excitations. This is achieved at
$I_{\text{HeNe}} =$ 1.82 W/cm$^2$ as displayed in Fig.~\ref{fig1}(d)
\cite{noteN0}.
The $N=1$ regime is characterized by a single excitation 
at $2\hbar \omega_0$
(degenerate in spin and charge) between FD levels with the same angular
momentum. Therefore its energy
position measures directly $\hbar \omega_0 = 1.1$ meV,
as shown in Fig.~\ref{fig1}(c). This condition is
reached at $I_{\text{HeNe}} = 1.46$ W/cm$^2$.
Peculiar to this regime is the absence of
multiplets at higher energy, as confirmed by calculation of the
light scattering cross section within a perturbative approach
under the resonant conditions employed here \cite{Schuller99}.
\par
\begin{figure}
\begin{center}
\includegraphics*[width=9.1cm]{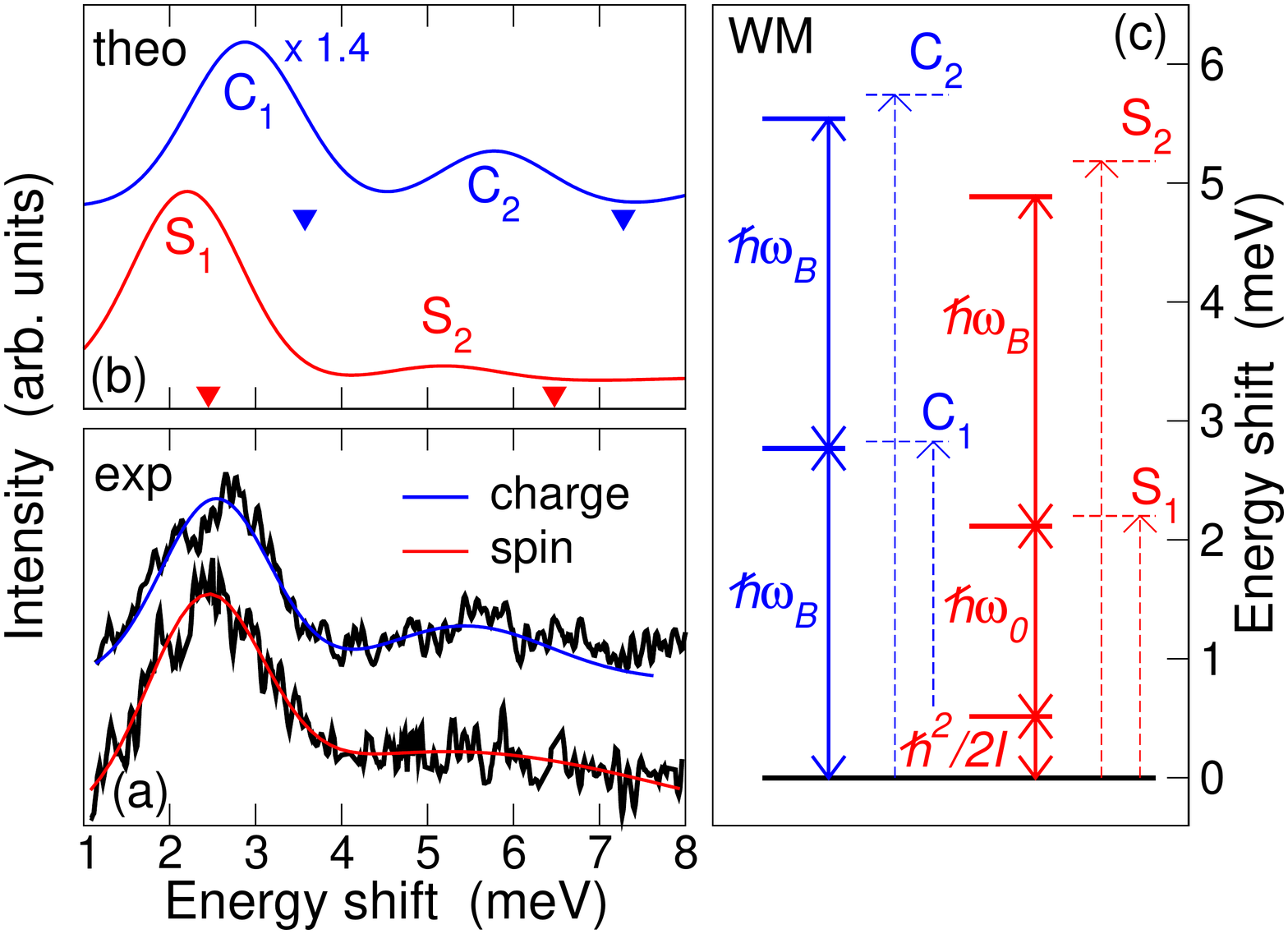}
\end{center}
\caption{(Color)  (a) Experimental spectra (black lines) of
spin and charge excitations at $I_{\text{HeNe}}=0.68$ W/cm$^{-2}$.
Red and blue lines are gaussian fits. (b) CI spectra
for $N=2$ (the peaks are artificially
broadened by gaussians with FWHM = 1.5 meV). The down triangles point
to the HF modes
for $\hbar\omega_0=$ 2.154 meV. (c) Comparison between CI (dashed lines)
and WM modes (solid lines) with $\hbar\omega_0=$ 1.6 meV.
\label{fig2} }
\end{figure}

Additional spin and charge modes appear at higher
energy [see Figs.~\ref{fig2}(a)
and \ref{fig4}] as we further reduce $I_{\text{HeNe}}$.
The emergence of multiplets in the spectra experimentally defines 
the occurrence of two electrons in the QDs at
$I_{\text{HeNe}} = 0.68$ W/cm$^2$.
In this regime of shallow confinement,
$\hbar \omega_0 \approx$ 1--2 meV, the electrons are
highly correlated \cite{egger, reimann, kalliakos08}.
To identify the degree of
correlation of the two electrons we theoretically evaluate the spectra
employing the full CI method \cite{Rontani2006}.
This  provides accurate
energies and wave functions of strongly interacting systems, by
bulding $N$-body states as linear combinations of many Slater
determinants. The latter are obtained by filling with $N$ electrons
in all possible ways a truncated FD level basis set (here
truncation includes the lowest 10 shells, with a relative error on
excitation energies of $\approx 10^{-4}$). From CI wave functions of
ground and excited $N$-body states
we evaluate inelastic light scattering
matrix elements and spectra \cite{kalliakos08, garcia}.
\par
The spectra predicted by CI calculations for $N=2$ [Fig.~\ref{fig2}(b)]
are remarkably similar to the experimental ones. In the evaluations 
the only free parameter, $\hbar\omega_0$,
is adjusted to fit the energy of the four peaks shown
in Fig.~\ref{fig2}(a). The spectra appear as two doublets ---each formed
by two peaks in the charge (blue lines) and spin (red lines)
channels, labeled respectively as C$_1$, S$_1$ and C$_2$, S$_2$ in 
Fig.~\ref{fig2}(b). Both spin excitations of Fig.~\ref{fig2}(a) 
are slightly red shifted ($\approx \ $0.2 meV)
from their charge counterparts. 
These splittings are also seen in the theoretical calculations
in Fig.~\ref{fig2}(b), which also reproduce the relative intensity of the peaks
in each channel. The value of $\hbar\omega_0=$ 1.6 meV used
in Fig.~\ref{fig2}(b) is larger than that measured
at $N=1$, as expected for a lower $I_{\text{HeNe}}$ \cite{kalliakos08nano}.
\par
The down triangles in Fig.~\ref{fig2}(b) summarize instead the results of a
self-consistent Hartree-Fock (HF) calculation \cite{Rontani99} of spin
and charge transition energies that neglects correlation effects. This
approach was extensively used to interpret previous light scattering spectra
in the many-electron regime \cite{Strenz94, Lockwood96, Schuller98}.
By fitting the position of S$_1$ [lowest down triangle
in Fig.~\ref{fig2}(b)] to its measured value
[Fig.~\ref{fig2}(a)] by properly tuning
$\hbar\omega_0$, then the remaining three HF
excitations are blue shifted more than 1 meV from
their measured values. On the other hand, by fitting the position of the
higher HF doublet (not shown) then S$_1$ is red shifted by $\approx 1$ meV.
This analysis therefore rules out an interpretation
in terms of single-particle transitions between FD shells and 
establishes the onset of electron correlation
as we increase the electron population from $N=1$ to $N=2$.
\par
\par

The impact of Coulomb correlations is represented by the
dimensionless radius $r_s$ of the circle whose area is equal
to the area per electron (in units of the
Bohr radius \cite{garcia}). For the CI calculation of Fig.~\ref{fig2}(b)
$r_s=3.39$. In this dilute regime
we expect the two electrons to form a molecule.
This is confirmed in Fig.~\ref{fig2}(c) by comparing the CI excitations
of Fig.~\ref{fig2}(b) [dashed lines in Fig.~\ref{fig2}(c)] 
with the analytical predictions
for a perfectly formed ``Wigner'' molecule
\cite{matulis01} (WM) [solid lines in Fig.~\ref{fig2}(c)]. 
The WM excitations have a composite nature:
C$_1$ (C$_2$) is a single (double) excitation of the breathing mode,
with energy $\hbar\omega_B$ ($2\hbar\omega_B$), S$_1$ is a dipolar
center-of-mass excitation $\hbar\omega_0$ plus a rotational quantum
$\hbar^2/2I$ (the system as a whole keeps $M=0$), and S$_2$
adds a breathing mode quantum. 
Remarkably, spin and charge
excitations scale differently with $\hbar\omega_0$, since
$\omega_B=\sqrt{3}\omega_0$ whereas $I\propto \omega_0^{-4/3}$.
The WM estimate for $\hbar^2/2I$, and hence the singlet-triplet
splitting, is 0.517 meV.

\begin{figure}
\begin{center}
\includegraphics*[width=5.5cm]{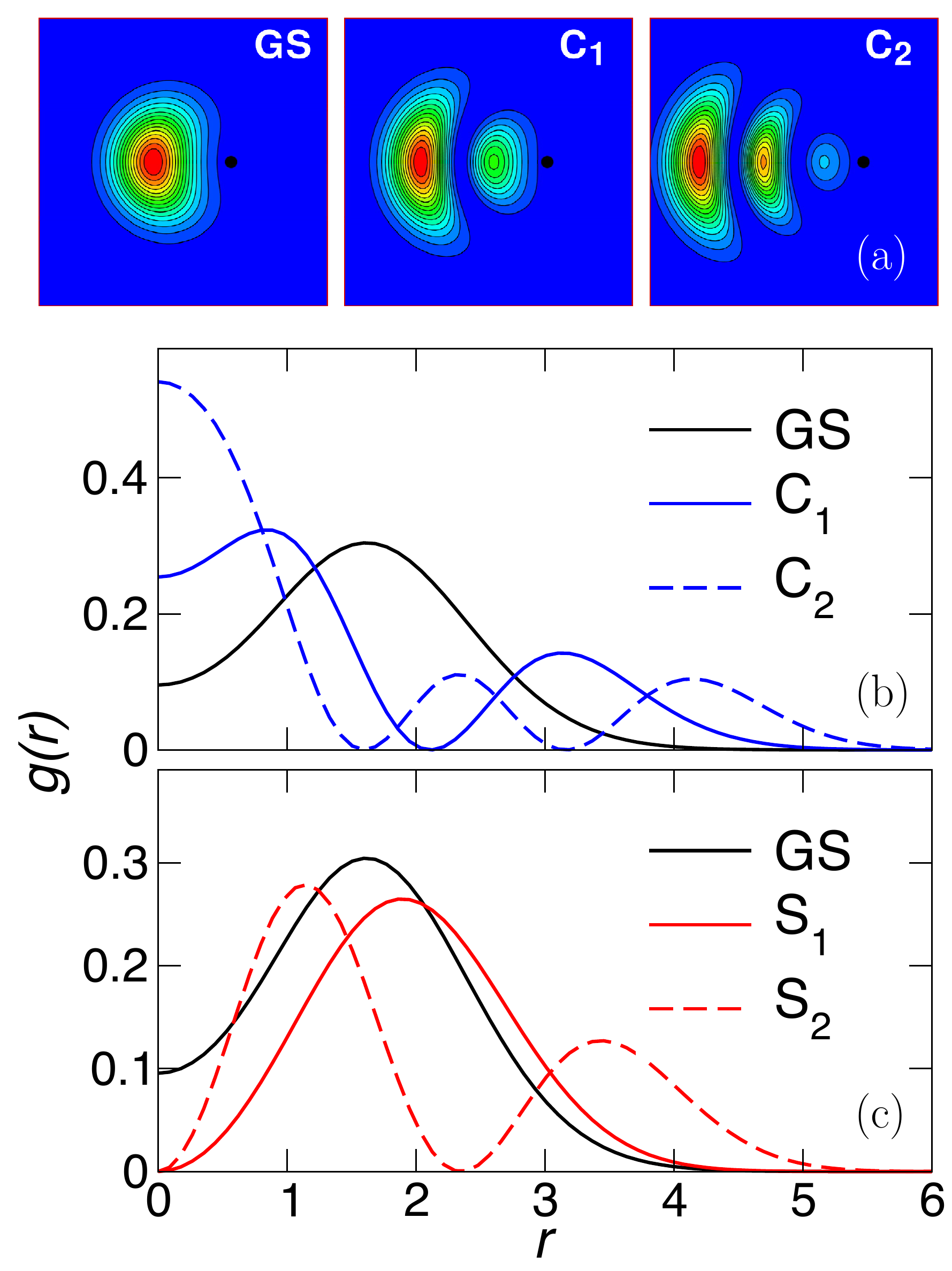}
\end{center}
\caption{(Color)  Pair correlation functions for the CI states assigned
in Fig.~\ref{fig2}.
The unit length is $\ell_{\text{QD}}=(\hbar/m^* \omega_0)^{1/2}$.
(a) Conditional probability $P(x,y;x_0,y_0)$ of measuring an electron in the
$xy$ plane provided another one is fixed at position $(x_0,y_0=0)$,
labelled by a black dot, where $x_0$ is located at the average value of
$r$. From left to right: ground (GS) and excited states C$_1$ and
C$_2$, respectively. The squares' size is $7\times 7$, and the
15 equally spaced contour levels go from blue (minimum) to red (maximum).
(b-c) Pair correlation functions $g(r)$ vs $r$ for the ground state,
charge (b), and spin (c) excitations, respectively. \label{fig3}}
\end{figure}

\begin{figure}
\begin{center}
\includegraphics[width=5.5cm]{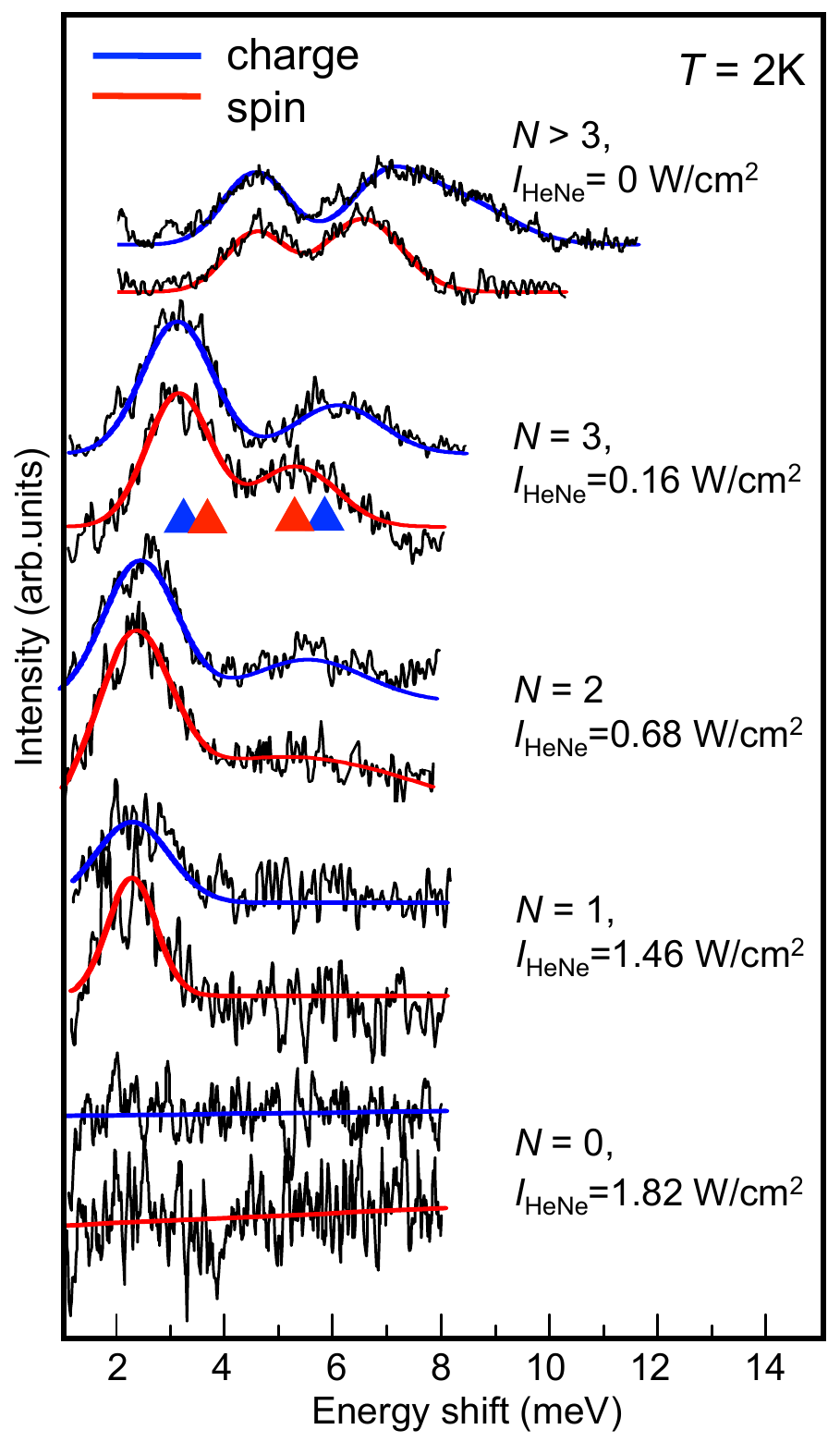}
\end{center}
\caption{(Color)  Evolution of spin and charge excitations
as a function of $I_{\text{HeNe}}$. Curves
are best-fits to the data with zero, one or two gaussians.
Up triangles point to the CI energies of spin (red) and
charge (blue) modes. \label{fig4}}
\end{figure}

The agreement between CI and WM
predictions in Fig.~\ref{fig2}(c) is very good
---with small deviations only for the
higher excitations--- implying that electrons in the ground state
freeze their relative distance   
(cf.~Fig.~\ref{fig3} and Supplementary Discussion). 
The formation of this Wigner molecule, which develops after the emergence of 
rotational bands \cite{kalliakos08} as $r_s$ increases, occurs 
relatively soon for the dimer since it involves only nearest-neighbor
particles (cf.~Supplementary Discussion).

\par

In Fig.~\ref{fig3}(a) we plot the conditional probability
$P(x,y;x_0,y_0)$ of finding an electron at position $(x,y)$
provided the other one is fixed at $(x_0,y_0)$
(black dot). In the ground state (GS, left panel) the well
developed correlation hole around the fixed electron suggests
the freezing of the relative distance $r$.
This is best seen from the pair correlation
function $g(r)$ [black line in Fig.~\ref{fig3}(b)], which for $N=2$ is
the square modulus of the relative-motion wave function.
Since $g(r)$ is the probability that two electrons are at
distance $r$, its well developed maximum around $r\approx 1.5$
[in units of the characteristic QD lenght $\ell_{\text{QD}}=
(\hbar/m^*\omega_0)^{1/2}$] identifies the inter-electron equilibrium
distance of the molecule. Besides, the plots of C$_1$ and C$_2$
in Fig.~\ref{fig3}(a) [Fig.~\ref{fig3}(b)] show respectively one
and two additional
nodal surfaces in the $xy$ plane (nodes along $r$), pointing to
the excitation of respectively one and two quanta of the breathing mode.
Note that the molecule is not perfectly formed due to
residual electron delocalization [$g(r=0)\neq 0$ in Fig.~\ref{fig3}(b)].
Spin excitations [red lines in Fig.~\ref{fig3}(c)] have a different nature:
$g(r=0)=0$ for symmetry [$g(r)\propto r^{M_{\text{rel}}}$
as $r\rightarrow 0$ and $M_{\text{rel}}=1$] and S$_1$ (S$_2$)
has no node (one node) at finite $r$, consistently with the
the excitation of no (one) breathing mode quantum. A similar
analysis in the center-of-mass coordinates confirms
the assignments of Fig.~\ref{fig2}(c).

\par
Figure \ref{fig4} summarizes the evolution of the measured excitations
as a function of $I_{\text{HeNe}}$. The 
data show that: (i) $\hbar \omega_0$ increases at lower $I_{\text{HeNe}}$
since the modes are blue shifted \cite{kalliakos08nano};
(ii) $N$ increases for $I_{\text{HeNe}} < 0.68$ W/cm$^2$ since 
the two low-lying spin
and charge modes do not split significantly, contrary 
to the expected behavior for $N=2$ \cite{CInote}.
Indeed the spectra at
$I_{\text{HeNe}} = 0.16$ W/cm$^2$ agree well with
the CI excitation energies for
$N=3$, shown as up triangles in Fig.~\ref{fig4}.
The best agreement with the experimental
data is obtained with $r_s=2.18$ ($\hbar\omega_0 = 2.8$ meV),
a value significantly smaller than the one
for $N=2$. This decrease of $r_s$ weakens
electron correlations and it stops at
$I_{\text{HeNe}}=0$. Then the spectra resemble those obtained
in similar QDs studied previously by us \cite{garcia, kalliakos08}
and interpreted as due to the low-lying spin and charge
excitations of $N>3$ electrons with
$r_s \approx 2$.
\par
In conclusion, we have reported the characteristic excitations of two
correlated electrons (a molecular dimer) in nanofabricated GaAs quantum dots.
Charge tunability with single electron accuracy is obtained by photodepletion.
These studies open new opportunities for understanding the ground states of
few electrons in semiconductor quantum dots in the regime of strong
Coulomb interaction and weak confinement potential.

\begin{acknowledgments}
V. P. and M. R. are supported by the projects MIUR-PRIN 2008H9ZAZR
and INFM-CINECA 2009. A. P. is supported by the NSF Nanoscale Science
and Engineering Initiative Award No. CHE-0641523, by the NSF Grants No.
DMR-0352738 and DMR-0803445, by the DOE Grant
No. DE-AIO2-04ER46133, and by a grant from the W. M. Keck
Foundation. We thank E. Molinari and G. Goldoni for helpful discussions.
\end{acknowledgments}

\section*{Supplementary Discussion: Theoretical analysis of
Wigner localisation in a broad range of $r_s$}

Since the electron dimer is a finite system, there is no clear-cut
transition to the WM state as $r_s$ increases. A good indicator of
the crossover to the WM exploits the irrelevance of the spin degree 
of freedom in this classical limit \cite{bryant, egger}.
Specifically, the pair correlation functions $g(r)$ for the ground
and S$_1$ states shown in Fig.~3(c) must overlap in the WM limit.
In fact, these represent the square moduli of the relative-motion
wave functions. Since in the WM limit S$_1$ and the ground state 
differ only for the excitation of one rotational rigid-body quantum 
plus a center-of-mass oscillation (cf.~main text), the relative 
motion ---and hence $g(r)$--- is left unchanged. However, the two states 
have different spin symmetries: S$_1$ and the ground state
are a spin triplet and singlet with one node and no
node for $g(r)$ at $r=0$, respectively. Therefore the $g(r)$'s
are generically different and overlap only in the WM limit,
when all the wave function weight is removed from the origin.
Consistently, electrons fully localise in space.

\begin{figure}
\begin{center}
\includegraphics*[width=8.9cm]{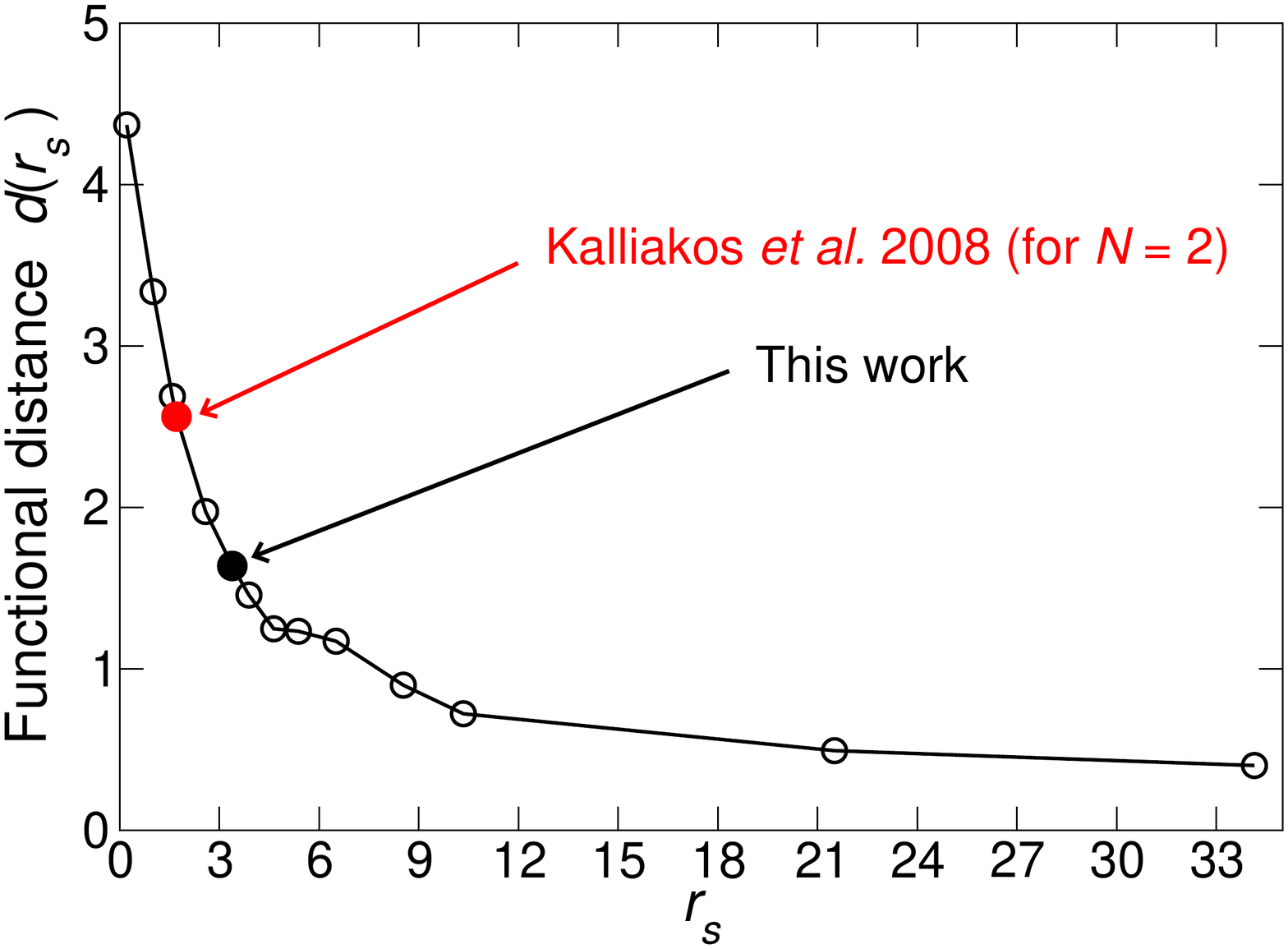}
\end{center}
\caption{
Functional distance $d(r_s)$ vs $r_s$ as obtained by
CI calculations.
The unit length is $\ell_{\text{QD}}=(\hbar/m^* \omega_0)^{1/2}$.
The continuous line is a guide to the eye.
\label{spindiff}}
\end{figure}

This crossover to the WM is analyzed quantitatively
by evaluating the functional distance $d(r_s)$
between the pair correlation
functions $g(r)$ of the ground state [$g_{\text{GS}}(r)$]
and S$_1$ [$g_{\text{S}_1}(r)$]:
\begin{equation}
d(r_s)=2\pi \int_0^{\infty}dr\,r\left|g_{\text{GS}}(r)
-g_{\text{S}_1}(r) \right|.
\end{equation}
$d(r_s)$ is plotted as a function of $r_s$ in Fig.~\ref{spindiff}.
Note that the evolution is smooth \cite{ghosal, kalliakos08},
with $d(r_s)\rightarrow 0$ only for $r_s\rightarrow \infty$.
Nevertheless, one may locate the crossover to the WM state at
$r_s\approx 4$ where the slope of $d(r_s)$ significantly 
changes. This estimate
is consistent with the theoretical analysis of Ref.~\onlinecite{egger}. 
Note the proximity of the experimental value (filled black circle
in Fig.~\ref{spindiff}) to the crossover region, which confirms our
interpretation in terms of the WM dimer.

In a previous work \cite{kalliakos08} we have demonstrated
the emergence of rotational bands for $N=4$ at relatively
high density ($r_s=1.71$). This transition to a correlated molecular state
is distinct from the formation of the WM, which implies electron 
localisation and occurs at larger values of $r_s$ \cite{notaWigner}. 
This may be seen in Fig.~\ref{spindiff}, where the 
value of $d(r_s)$ for $N=2$ corresponding to the same density as in
Ref.~\onlinecite{kalliakos08} (red filled circle in
Fig.~\ref{spindiff}) is significantly distant from the crossover region.

\begin{figure}
\begin{center}
\includegraphics*[width=9cm]{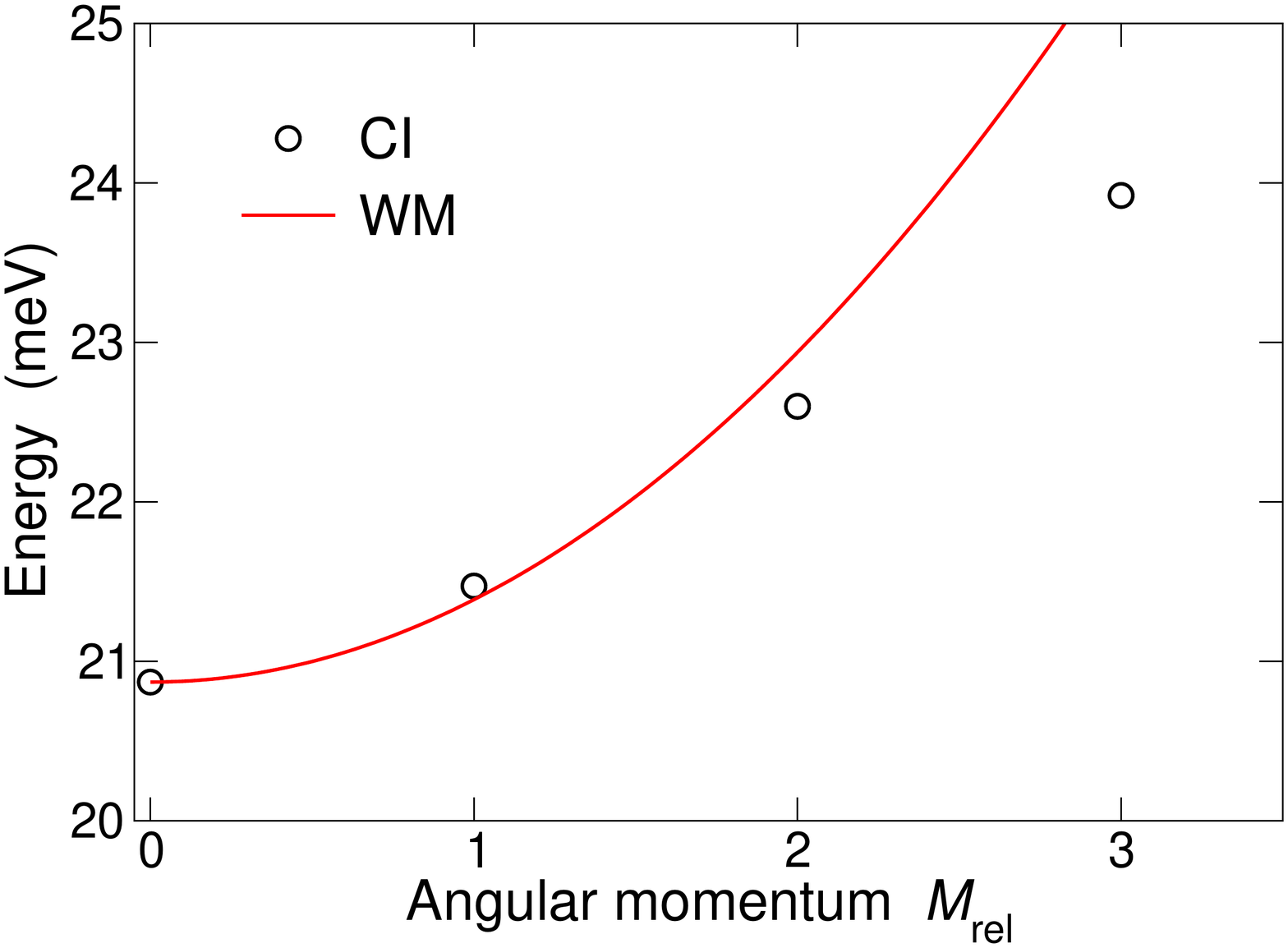}
\end{center}
\caption{CI lowest-state energy of
the electron dimer vs relative angular momentum $M_{\text{rel}}$
for $r_s = 3.39$. The continuous red line is the lowest
rotational band of a perfectly formed WM
(with an arbitrary offset).
\label{rotational} }
\end{figure}

The assignment of rotational bands in Ref.~\onlinecite{kalliakos08} 
identifies states with different values of 
$M_{\text{rel}}$ and same profiles of $g(r)$. A feature peculiar 
of the WM is the parabolic dispersion of the rotational band,
$ \sim \hbar^2M^2_{\text{rel}}/ 2I$,
with the momentum of inertia $I$ univocally fixed by $r_s$.  
This may be seen in Fig.~\ref{rotational}, where we compare
the CI lowest-energies of the dimer for each $M_{\text{rel}}$ sector
(black circles) with the parameter-free WM rotational band
(continuous red line). The agreement is satisfactory up to $M_{\text{rel}}=2$
and then deteriorates as $M_{\text{rel}}$ increases. 
The rationale is that the WM melts as the energy ($M_{\text{rel}}$)
increases, the larger $r_s$ the higher the energy.
According to Ref.~\onlinecite{yannouleas00}, the
results of Fig.~\ref{rotational}
show that the WM is ``floppy'' for this value of $r_s$.


\begin{thebibliography}{40}
\bibitem{noteN0} A small increase of the signal can be seen in the spin 
channel at around 2 meV. This might indicate that a few QDs are already 
populated by one electron.
\bibitem{CInote} 
The HF splitting for $N=2$ is an exchange energy term, which  
increases with $\hbar\omega_0$ like $(\hbar\omega_0)^{1/2}$.
Such increase is confirmed by the full CI simulations.
\bibitem{notaWigner} The criterion we have used in
Ref.~\onlinecite{kalliakos08} to investigate the formation of the WM
(cf.~Fig.~4c in \cite{kalliakos08}) relies on the
spin polarization of the four-electron ground state and
therefore it is not applicable to the unpolarized dimer ground state.
\end{thebibliography}
\end{document}